\documentclass[a4paper,fleqn,usenatbib]{mnras}
\usepackage[T1]{fontenc}
\usepackage{ae,aecompl}

\usepackage{graphicx}
\usepackage{txfonts}
\usepackage{comment}

\title[The X-ray spectral complexity in NLS1s]{A correlation between [O III] line property and X-ray spectral complexity in narrow-line Seyfert 1 galaxies?}

\author[S. Chen et al.]{
S. Chen, $^{1,2,3}$ \thanks{E-mail: sina.chen@phd.unipd.it}
G. La Mura, $^{4}$
M. Berton, $^{5,6}$
L. Foschini, $^{7}$
E. Congiu, $^{8,7}$
\newauthor
F. Di Mille, $^{8}$
S. Ciroi, $^{2,9}$
E. Bottacini, $^{2,10}$
J.H. Fan, $^{1}$
and A. Vietri $^{2}$
\\
$^{1}$ Center for Astrophysics, Guangzhou University, 510006, Guangzhou, China \\
$^{2}$ Dipartimento di Fisica e Astronomia "G. Galilei", Universit{\`a} di Padova, Padova, Italy \\
$^{3}$ Istituto Nazionale di Fisica Nucleare (INFN), Sezione di Padova, 35131, Padova, Italy \\
$^{4}$ LIP - Laboratory of Instrumentation and Experimental Particle Physics, Av. Prof. Gama Pinto 2, 1649-003 Lisboa, Portugal \\
$^{5}$ Finnish Centre for Astronomy with ESO (FINCA), University of Turku, Quantum, Vesilinnantie 5, 20014 University of Turku, Finland \\
$^{6}$ Aalto University Mets{\"a}hovi Radio Observatory, Mets{\"a}hovintie 114, FIN-02540 Kylm{\"a}l{\"a}, Finland \\
$^{7}$ INAF - Osservatorio Astronomico di Brera, Via E. Bianchi 46, 23807, Merate (LC), Italy \\
$^{8}$ Las Campanas Observatory, Carnegie Institution for Science, Colina El Pino Casilla 601, La Serena, Chile \\
$^{9}$ INAF - Osservatorio Astronomico di Padova, Vicolo dell'Osservatorio 5, 35122 Padova, Italy \\
$^{10}$ W.W. Hansen Experimental Physics Laboratory \& Kavli Institute for Particle Astrophysics and Cosmology, Stanford University, USA
}

\date{Accepted XXX. Received YYY; in original form ZZZ}

\pubyear{2019}

\begin{document}
\label{firstpage}
\pagerange{\pageref{firstpage}--\pageref{lastpage}}
\maketitle

\begin{abstract}
We present a detailed study of 11 narrow-line Seyfert 1 galaxies (NLS1s) from the Six-degree Field Galaxy Survey (6dFGS) that both have optical and X-ray spectroscopic observations. There are five complex NLS1s (C-NLS1s) and six simple NLS1s (S-NLS1s). We propose a possible correlation between [O III] line asymmetry and X-ray complexity. The outflow or wind from the inner accretion disk is commonly present in NLS1s and mostly directed along the system axis. In C-NLS1s only weak wind effects are measured, the X-ray spectral complexity might be caused by the presence of ionized material in the wind. On the contrary, the wind in S-NLS1s is fast, the ionized material could be swept by such a strong wind, thus the complex feature is missing which results in a simple X-ray spectrum. Furthermore, this outflow scenario seems to be an inclination effect. Since the speed of the wind is higher in a small inclination while lower in a large inclination, S-NLS1s might be sources viewed at small angles while C-NLS1s might be sources viewed at large angles.
\end{abstract}

\begin{keywords}
galaxies: active - galaxies: nuclei - galaxies: Seyfert - quasars: emission lines - quasars: supermassive black holes
\end{keywords}

\section{Introduction}

Narrow-line Seyfert 1 galaxies (NLS1s) are intriguing members in the active galactic nuclei (AGN) family. Compared to regular broad-line Seyfert 1 galaxies (BLS1s), they have narrow H$\beta$ line, weak [O III] lines, and strong Fe II multiplets in their optical spectra \citep{Osterbrock1985, Goodrich1989}. These optical properties suggest that NLS1s are powered by low-mass black holes, typically in the range of $M_{BH} \sim 10^6 - 10^8 M_{\odot}$, while the central black hole masses of BLS1s are of the order of $10^7 - 10^9 M_{\odot}$ \citep{Grupe2000, Jarvela2015, Cracco2016, Chen2018a}. NLS1s are characterized by a high accretion luminosity as well, usually close to the Eddington limit \citep{Boroson1992, Collin2004}. All of these indicate that NLS1s might be an early phase of AGN evolution \citep{Mathur2000}. The $\gamma$-ray emission detected by the \textit{Fermi} Large Area Telescope (LAT) in NLS1s \citep{Abdo2009a, Abdo2009b, Foschini2011, Foschini2015, Yao2015, Liao2015, D'Ammando2015, D'Ammando2016, Paliya2016, Berton2017, Paliya2018, Lahteenmaki2018, Yao2019} confirmed the presence of relativistic jets in this peculiar subclass of AGN. However, the central black hole masses of NLS1s are generally two orders of magnitude less than those of blazars ($M_{BH} \sim 10^8 - 10^{10} M_{\odot}$). Additionally, NLS1s are typically hosted in spiral galaxies \citep{Crenshaw2003, Deo2006, Orbandexivry2011, Kotilainen2016, Olguiniglesias2017}, while blazars are usually hosted in elliptical galaxies \citep{Sikora2007}. Currently it is still puzzling how the relativistic jets form in these young AGN, although merging and galaxy interaction are strong suspects \citep{Jarvela2018, Berton2019}.

It is well-known that AGN with high Eddington ratios, such as NLS1s, generate powerful outflows or winds by the radiation pressure from the accretion disk \citep{Proga2000, Tombesi2010}. These outflows are generally associated with the presence of asymmetric [O III] lines, which can be decomposed in two distinct components. The narrow component is the line core, typically having the same redshift as the whole galaxy. The broad component is the blue wing, usually connected with a gas outflow in the narrow-line region (NLR) and blueshifted with respect to their rest-frame wavelength \citep{Greene2005a}. In some cases, the entire line is blueshifted \citep{Zamanov2002, Marziani2003}. The mechanism of this [O III] shift is not well understood. A common hypothesis is that such feature is generated by the strong winds from the accretion disk \citep{Komossa2008}. Alternatively, a powerful relativistic jet can affect the gas kinematics in the NLR and induce a bulk motion in the NLR which translates into the observed blueshift \citep{Berton2016a}.

In the X-ray regime, NLS1s display extreme behaviors. They exhibit fast variability, typically on short time scale less than one day \citep{Boller2000}, as expected for the small black hole mass and high accretion rate \citep{Ponti2012}. In the soft X-ray band (below 2 keV), NLS1s show strong soft X-ray excess above the prediction of a single power-law \citep{Boller1996}. The origin of the soft X-ray excess is still unclear. An explanation is the low temperature optically thick Comptonization model, which includes an additional cool Comptonizing component from an inner disk producing the soft excess \citep{Gierlinski2004, Done2012}. Another one is the relativistically blurred photoionized reflection from an inner accretion disk, which reproduces the emission expected from an optically thick photoionized disk around a black hole \citep{Ross2005, Crummy2006}. Both models fit the spectra very well in the X-ray range, thus it is difficult to tell them apart. In the hard X-ray band (2-10 keV), NLS1s generally have steeper intrinsic spectra than BLS1s \citep{Brandt1997}. This might suggest that photons from the soft excess of NLS1s could Compton cool the coronae, which create lower temperature and more diffuse coronae than in BLS1s and thereby steepen the hard X-ray continua of NLS1s \citep{Maraschi1997}.

Some NLS1s exhibiting extremely complex features around Fe K-shell at 6-8 keV, are classified as complex NLS1s (C-NLS1s), while the others are described as simple NLS1s (S-NLS1s), whose 2-10 keV spectra do not strongly deviate from a single power-law continuum \citep{Gallo2006}. The same authors proposed that C-NLS1s are in a low X-ray flux state, whereas S-NLS1s are in a normal X-ray flux state. However, a research by \citet{Jin2017a, Jin2017b} proposes a global picture for the structure of a super-Eddington accretion flow where the inner disk puffs up, shielding much of the potential NLR materials, and appearing as an inclination effect with respect to the clumpy disk wind. In this picture, simple and complex NLS1s can be interpreted within the same accretion flow scenario but with different viewing angles. The inclination of S-NLS1s is low and the line of sight is unobscured showing original X-ray emission produced by the coronae. Instead, C-NLS1s are those seen at large viewing angles and the line of sight is intercepted by absorption materials.

Understanding the physical difference between simple and complex NLS1s would be a breakthrough in our understanding of the X-ray emission mechanism of this peculiar class of AGN. Recently, \citet{Chen2018a} presented a new catalog of NLS1s in the southern hemisphere from the Six-degree Field Galaxy Survey (6dFGS) final data release (DR3) \citep{Jones2009} \footnote{http://www-wfau.roe.ac.uk/6dFGS/.}. They classified 167 sources as NLS1s based on their optical spectral properties. In this paper, we selected sources both having optical spectroscopic observations and X-ray observations by the \textit{XMM-Newton} with public data on the \textit{XMM-Newton} Science Archive (XSA). Our aim is to investigate whether a link is present between the optical and X-ray spectra, in particular between [O III] line properties and X-ray complexity. Throughout this work, we adopt a standard $\Lambda$CDM cosmology with a Hubble constant $H_0$ = 70 km s$^{-1}$ Mpc$^{-1}$, $\Omega_{\Lambda}$ = 0.73 and $\Omega_{M}$ = 0.27 \citep{Komatsu2011}. We assume the flux and spectral index convention $ F_{\nu} \propto \nu^{-\alpha} $.

\section{Sample selection and data reduction}

\subsection{Sample selection}

We searched the 6dFGS NLS1 sample \citep{Chen2018a} in the XSA and selected objects having an EPIC \footnote{The European Photon Imaging Camera.} pointed observation within a search radius of 10 arcmin. We excluded some sources with a total counts of PN camera less than 3000. If the X-ray spectrum is faint, the complex features are likely not visible, leading to a misclassification of the source as a simple one. In total, we found 11 NLS1s that can be used in this work. 

\subsection{Optical data reduction}

We collected the optical spectra for these 11 NLS1s. Nine objects have new observations by the IMACS instrument of the 6.5-meter Walter Baade telescope and the WFCCD instrument of the 2.5-meter Du Pont telescope at the Las Campanas Observatory (LCO) in Chile. We performed a standard spectroscopic reduction using IRAF \footnote{http://iraf.noao.edu/}, with bias and flat-field correction, followed by wavelength, flux calibration, and sky subtraction. Standard stars EG21, LTT2415, and LTT7987 were used for the flux calibration. The spectra of two more objects were obtained from the archive of the 2.54-meter Isaac Newton telescope in La Palma (Spain) and the 1.22-meter Galileo telescope in Asiago (Italy). All these spectra were corrected for galactic extinction using the $A_V$ extinction coefficients (Landolt V) \citep{Schlafly2011} extracted from the NASA/IPAC \footnote{The National Aeronautics and Space Administration / The Infrared Processing and Analysis Center.} Extragalactic Database (NED) \footnote{https://ned.ipac.caltech.edu/.}. For the redshift correction, we used the low ionization forbidden lines, such as [O I] $\lambda$6300, [S II] $\lambda\lambda$6716,6731, and [O II] $\lambda$3727 as reference \citep{Komossa2008,Berton2016a}. If these lines were not visible, we brought the optical spectra to the rest frame according to the redshift in the NED. The spectroscopic observation details are listed in Table~\ref{optical}.

\begin{table}
\centering
\caption{The spectroscopic observation details of the 11 NLS1s.}
\label{optical}
\scriptsize
\begin{tabular}{ccccc}
\hline
\hline
6dFGS name & Short name & Telescope & Observed date & Exp. time \\
- & - & - & (yyyy-mm) & (s) \\
\hline
6dF J0228152-405715 & J0228 & Du Pont & 2019-02 & 3 $\times$ 900 \\
6dF J0230055-085953 & J0230 & Isaac Newton & 1996-08 & 2 $\times$ 900 \\
6dF J0436223-102234 & J0436 & Galileo & 2017-11 & 4 $\times$ 1200 \\
6dF J0452301-295335 & J0452 & Du Pont & 2019-02 & 3 $\times$ 900 \\
6dF J0708415-493306 & J0708 & Walter Baade & 2019-01 & 3 $\times$ 900 \\
6dF J1325194-382453 & J1325 & Walter Baade & 2019-01 & 1 $\times$ 600 \\
6dF J1511598-211902 & J1511 & Du Pont & 2018-07 & 3 $\times$ 900 \\
6dF J1638309-205525 & J1638 & Du Pont & 2018-07 & 3 $\times$ 600 \\
6dF J1937330-061305 & J1937 & Du Pont & 2018-07 & 3 $\times$ 900 \\
6dF J2135295-623007 & J2135 & Du Pont & 2018-07 & 3 $\times$ 900 \\
6dF J2245203-465211 & J2245 & Du Pont & 2018-07 & 3 $\times$ 900 \\
\hline
\end{tabular}
\end{table}

\subsection{X-ray data reduction}

The observation data files (ODFs) of these 11 NLS1s were processed with the \textit{XMM-Newton} Science Analysis System (SAS) \footnote{https://www.cosmos.esa.int/web/xmm-newton/sas.} version 16.1.0 with the latest current calibration files \footnote{https://www.cosmos.esa.int/web/xmm-newton/current-calibration-files.}. The PN camera was operated in full-frame mode, and the two MOS cameras were in large-window mode. We followed the standard prescriptions of the PN and MOS spectral extraction from point-like sources \footnote{https://www.cosmos.esa.int/web/xmm-newton/sas-threads.}. Spectra were extracted from the PN+MOS1+MOS2 CCDs. We filtered event lists for flaring particle background, and chose time intervals with low and steady background during the observations. Source photons were extracted in a circle with a radius of 30 arcsec centered on the source from the PN+MOS1+MOS2 images, and background was selected from a source-free region with a radius of 60 arcsec. For each source, ancillary response (arf) and redistribution matrix (rmf) files were created. Spectra were then rebinned to have at least 25 counts per bin. The \textit{XMM-Newton} observation details of these 11 NLS1s are listed in Table~\ref{xmm}.

\begin{table*}
\centering
\caption{The \textit{XMM-Newton} observation details of the 11 NLS1s.}
\label{xmm}
\footnotesize
\begin{tabular}{ccccccccc}
\hline
\hline
3XMM name & Short name & Counterpart & R.A. & Dec. & Observed date & CCD & Count rate & Exp. time \\
- & - & - & (hh:mm:ss) & (dd:mm:ss) & (yyyy-mm) & - & ($cts \, s^{-1}$) & (sec) \\
\hline
3XMM J022815.2-405714 & J0228 & IRAS 02262-4110 & 02:28:14.99 & -40:57:16.0	& 2004-12 & PN & 1.790 $\pm$ 0.008 & 26670 \\
&&&&&& MOS1 & 0.343 $\pm$ 0.003 & 31520 \\
&&&&&& MOS2 & 0.362 $\pm$ 0.003 & 31550 \\
\hline
3XMM J023005.5-085953 & J0230 & Mrk 1044 & 02:30:05.50 & -08:59:53.0 & 2013-01 & PN & 18.740 $\pm$ 0.015 & 85550 \\
&&&&&& MOS1 & 3.990 $\pm$ 0.007 & 84900 \\
&&&&&& MOS2 & 3.941 $\pm$ 0.007 & 82980 \\
\hline
3XMM J043622.2-102233 & J0436 & Mrk 618 & 04:36:22.24 & -10:22:33.8 & 2006-02 & PN & 9.733 $\pm$ 0.053 & 3538 \\
&&&&&& MOS1 & 2.149 $\pm$ 0.015 & 9630 \\
&&&&&& MOS2 & 2.515 $\pm$ 0.017 & 8883 \\
\hline
3XMM J045230.0-295335 & J0452 & IRAS 04505-2958 & 04:52:30.08 & -29:53:35.3 & 2003-09 & PN & 3.822 $\pm$ 0.021 & 8664 \\
&&&&&& MOS1 & 0.781 $\pm$ 0.008 & 13820 \\
&&&&&& MOS2 & 0.803 $\pm$ 0.008 & 13420 \\
\hline
3XMM J070841.4-493306 & J0708 & 1H 0707-495 & 07:08:41.49 & -49:33:06.0 & 2010-09 & PN & 5.466 $\pm$ 0.007 & 97710 \\
&&&&&& MOS1 & 1.111 $\pm$ 0.003 & 114000 \\
&&&&&& MOS2 & 1.079 $\pm$ 0.003 & 116400 \\
\hline
3XMM J132519.3-382452 & J1325 & IRAS 13224-3809 & 13:25:19.40 & -38:24:53.0 & 2016-08 & PN & 5.798 $\pm$ 0.007 & 108100 \\
&&&&&& MOS1 & 1.095 $\pm$ 0.003 & 125600 \\
&&&&&& MOS2 & 1.088 $\pm$ 0.003 & 124800 \\
\hline
3XMM J151159.8-211902 & J1511 & IRAS 15091-2107 & 15:11:59.78 & -21:19:01.6 & 2005-07 & PN & 2.895 $\pm$ 0.023 & 5650 \\
&&&&&& MOS1 & 0.906 $\pm$ 0.008 & 15090 \\
&&&&&& MOS2 & 0.910 $\pm$ 0.008 & 14190 \\
\hline
3XMM J163830.8-205524 & J1638 & IGR J16385-2057 & 16:38:31.09 & -20:55:25.0 & 2010-02 & PN & 2.641 $\pm$ 0.015 & 11490 \\
&&&&&& MOS1 & 0.761 $\pm$ 0.006 & 24080 \\
&&&&&& MOS2 & 0.759 $\pm$ 0.006 & 24080 \\
\hline
3XMM J193733.0-061304 & J1937 & IGR J19378-0617 & 19:37:33.01 & -06:13:04.8 & 2015-10 & PN & 19.970 $\pm$ 0.015 & 92900 \\
&&&&&& MOS1 & 4.534 $\pm$ 0.006 & 124600 \\
&&&&&& MOS2 & 5.041 $\pm$ 0.006 & 125300 \\
\hline
3XMM J213529.4-623006 & J2135 & IRAS F21325-6237 & 21:36:23.11 & -62:24:00.6 & 2006-11 & PN & 7.609 $\pm$ 0.121 & 525 \\
&&&&&& MOS1 & 1.886 $\pm$ 0.013 & 12060 \\
&&&&&& MOS2 & 1.902 $\pm$ 0.013 & 11580 \\
\hline
3XMM J224520.2-465211 & J2245 & IRAS F22423-4707 & 22:45:20.19 & -46:52:13.4 & 2008-11 & PN & 2.909 $\pm$ 0.036 & 2255 \\
&&&&&& MOS1 & 0.540 $\pm$ 0.010 & 5216 \\
&&&&&& MOS2 & 0.528 $\pm$ 0.010 & 5125 \\
\hline
\end{tabular}
\end{table*}

\section{Spectral analysis}

\subsection{Optical spectra}

\begin{figure}
\centering
\includegraphics[width=\columnwidth]{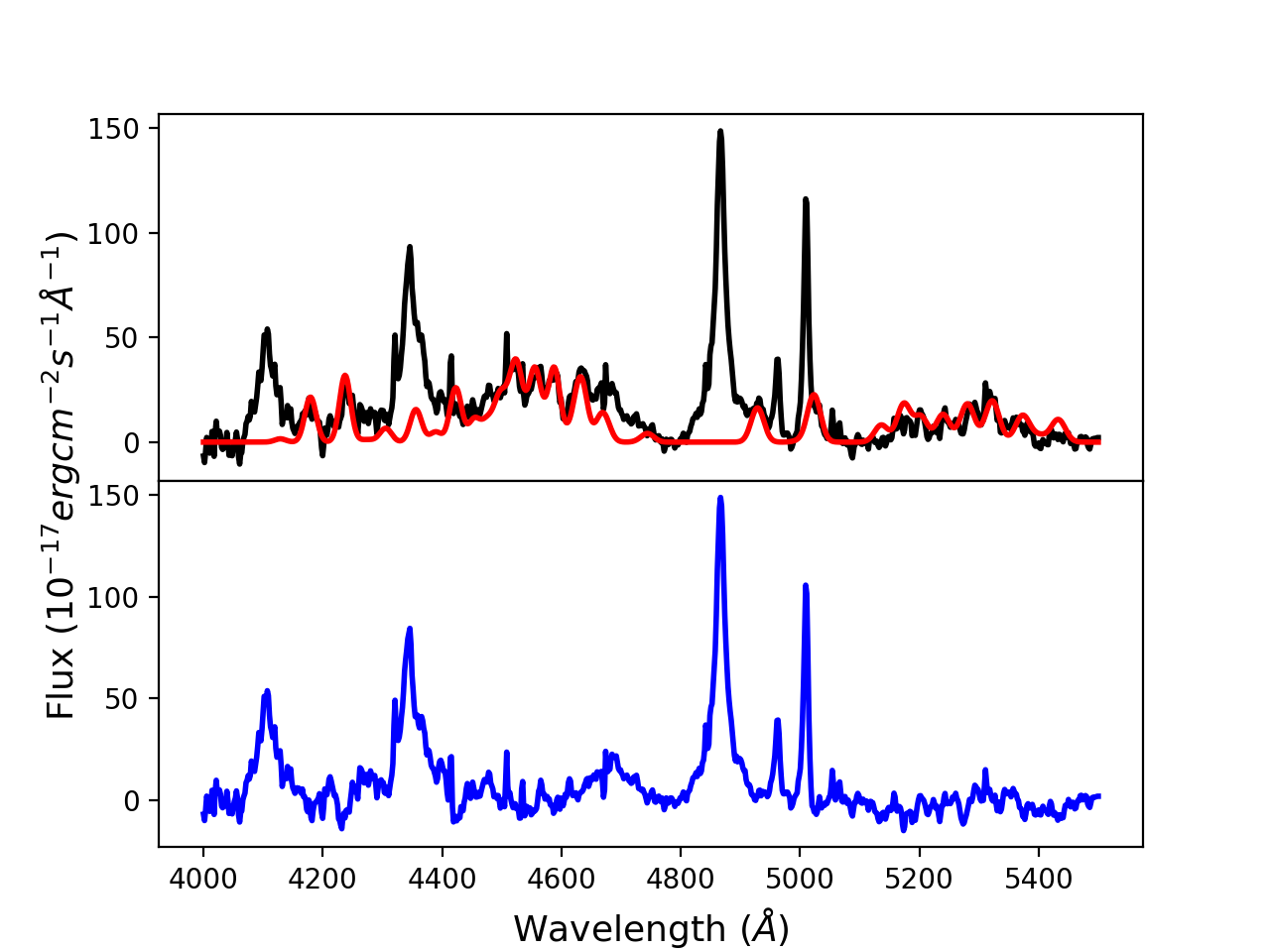}
\caption{The Fe II subtraction of IRAS F21325-6237. \textit{Top panel:} the observed spectrum with continuum subtracted (black line) and the Fe II model (red line). \textit{Bottom panel:} the spectrum with continuum and Fe II subtracted (blue line).}
\label{feii}
\end{figure}

\begin{figure}
\centering
\includegraphics[width=\columnwidth]{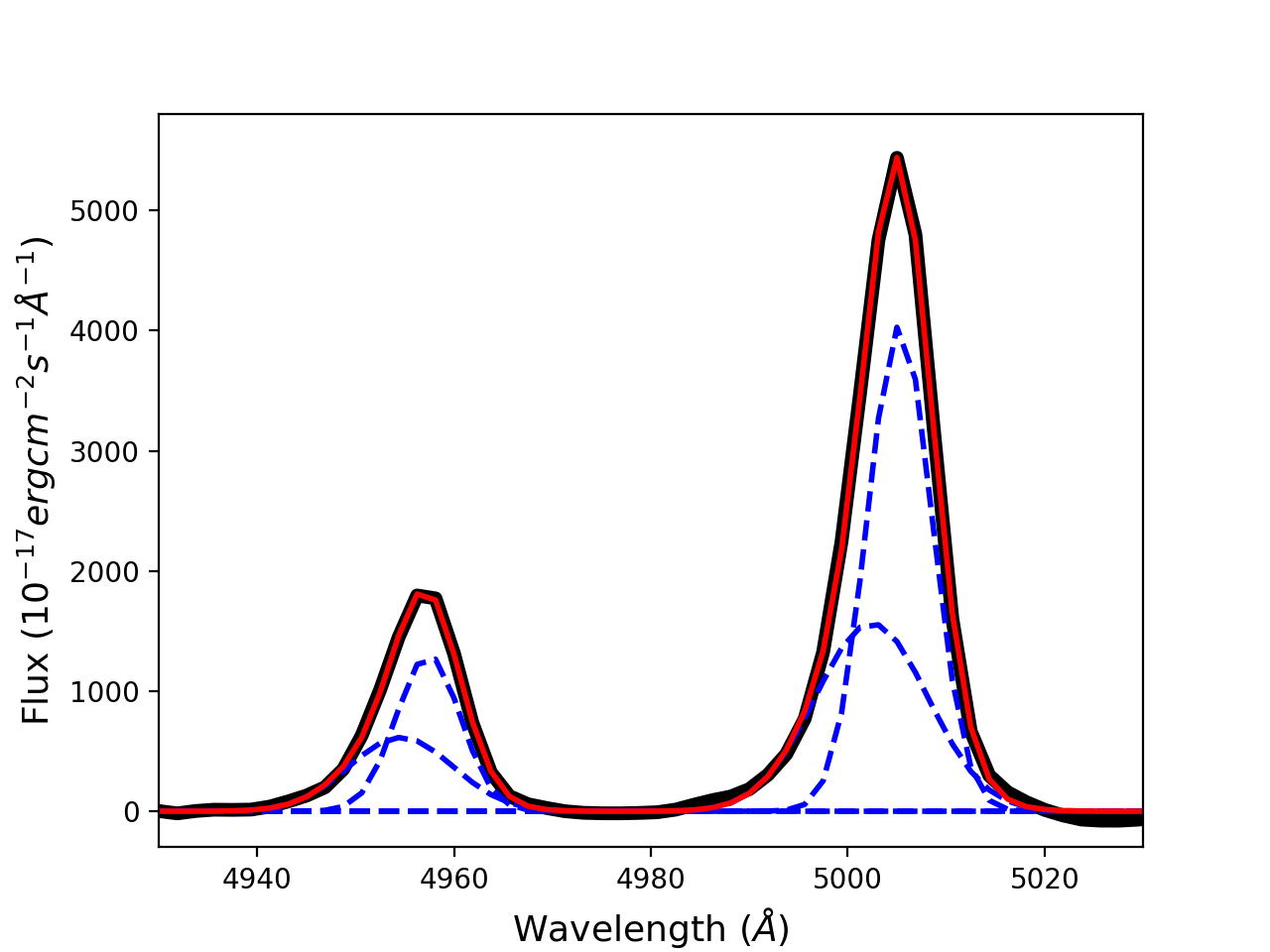}
\caption{The [O III] lines fitting of IGR J19378-0617. The black line indicates the observed spectrum with continuum and Fe II subtracted. The red line indicates the sum of the resulting fit. The dashed blue line indicates the Gaussian components.}
\label{oiii}
\end{figure}

The new optical spectra of the 11 NLS1s have a higher signal-to-noise (S/N) ratio than those from the 6dFGS, thus allowed us to analyze their emission line profiles in detail. We subtracted the continuum and the Fe II multiplets from the observed spectra. The Fe II templates were reproduced using the procedure described in \citet{Kovacevic2010, Shapovalova2012} \footnote{http://servo.aob.rs/FeII$\_$AGN/}. An example of Fe II subtraction is shown in Fig.~\ref{feii}. After these subtractions, we fitted the [O III] $\lambda\lambda$4959,5007 double lines with two Gaussian (one narrow and one broad) components and corrected for the instrumental resolution. An example of the [O III] lines fitting is displayed in Fig.~\ref{oiii}. The narrow and broad Gaussian components are associated with the line core and the blue wing, respectively.

We measured the velocity of the blue wing using the central wavelength of narrow component which represents the line core, minus that of broad component which represents the blue wing, as listed in Table~\ref{wavelength}. Although the fitting procedure has been carried out on both [O III] lines, we measured the velocity of the blue wing only for [O III] $\lambda$5007. The velocity measurements of [O III] $\lambda$5007 indeed should be more precise than those of [O III] $\lambda$4959, due to the weak intensity of [O III] $\lambda$4959 compared to that of [O III] $\lambda$5007. We also note that three sources do not have a wing velocity measurement. The [O III] lines in J0228 and J0708 are not detected, as common among sources with strong Fe II \citep{Marziani2018}. In J1325 instead the [O III] $\lambda$5007 line can be fitted with only one Gaussian component, possibly because of the poor S/N of the optical spectrum.

The errors were evaluated using a Monte Carlo method. We varied the line profile by adding a random Gaussian noise proportional to the root-mean-square (RMS) measured in the continuum. We then fitted the [O III] line with two Gaussian components as previously described, measuring the central wavelength and velocity again, and repeating the same process 100 times to estimate the standard deviation of each measurement. In this way, we obtained 1$\sigma$ errors for these parameters.

\begin{table*}
\centering
\caption{The central wavelength of [O III]$\lambda$5007 line components and the velocity of blue wing measurements. \textbf{Columns:} (1) short name, (2) redshift, (3) central wavelength of the line core, (4) central wavelength of the blue wing, (5) velocity of the blue wing, (6) X-ray classification, complex (C) or simple (S).}
\label{wavelength}
\begin{tabular}{cccccc}
\hline
\hline
Short name & Redshift & Line core & Blue wing & Velocity & Type \\
- & - & (\AA) & (\AA) & (km s$^{-1}$) & (C/S) \\
\hline
J0228 & 0.493 & - & - & - & S \\
J0230 & 0.016 & 5006.84 $\pm$ 1.66 & 5003.07 $\pm$ 2.41 & 226.06 $\pm$ 175.30 & C \\
J0436 & 0.036 & 5007.85 $\pm$ 1.16 & 5002.22 $\pm$ 5.42 & 336.64 $\pm$ 331.73 & S \\
J0452 & 0.247 & 5007.06 $\pm$ 0.05 & 4999.51 $\pm$ 0.68 & 452.12 $\pm$ 40.74 & S \\
J0708 & 0.041 & - & - & - & C \\
J1325 & 0.066 & - & - & - & C \\
J1511 & 0.045 & 5007.14 $\pm$ 0.96 & 5003.78 $\pm$ 1.37 & 201.02 $\pm$ 100.47 & S \\
J1638 & 0.026 & 5006.57 $\pm$ 2.09 & 5002.30 $\pm$ 3.37 & 255.51 $\pm$ 237.55 & C \\
J1937 & 0.010 & 5005.28 $\pm$ 1.26 & 5002.42 $\pm$ 2.59 & 171.08 $\pm$ 172.31 & C \\
J2135 & 0.061 & 5010.50 $\pm$ 2.28 & 5004.51 $\pm$ 2.90 & 358.41 $\pm$ 220.80 & S \\
J2245 & 0.200 & 5007.62 $\pm$ 0.02 & 4998.07 $\pm$ 0.23 & 571.65 $\pm$ 13.58 & S \\
\hline
\end{tabular}
\end{table*}

\subsection{X-ray spectra}

We modeled the PN, MOS1, and MOS2 spectra simultaneously using the XSPEC version 12.10.0 \citep{Arnaud1996} \footnote{https://heasarc.nasa.gov/xanadu/xspec/manual/node1.html.}. To identify the simple and complex NLS1s, we fitted the 2-12 keV spectra with a baseline model composing of a power-law continuum modified by Galactic absorption \citep{Gallo2006}. Sources whose 2-12 keV spectra do not strongly deviate from a simple power-law continuum with a null hypothesis of the baseline model greater than 0.10 are marked as S-NLS1s. Instead, sources that exhibit high-energy complexity with a null hypothesis of the baseline model less than 0.10 are marked as C-NLS1s. In total, there are six S-NLS1s and five C-NLS1s as listed in Table~\ref{classify}. However, it is worth noting that some objects might be misclassified due to the limited counts in the X-ray spectra.

\begin{table}
\centering
\caption{The X-ray spectral classification, reduced Chi-squared and null hypothesis of the baseline model fitting. \textbf{Columns:} (1) short name, (2) X-ray classification, complex (C) or simple (S), (3) reduced Chi-squared of the baseline model fitting, (4) null hypothesis of the baseline model fitting.}
\label{classify}
\begin{tabular}{cccc}
\hline
\hline
Short name & Type & $\chi^2_{\nu}$ & Null hypothesis \\
\hline
J0228 & S & 1.05 & 0.315 \\
J0230 & C & 1.76 & $<$ 0.10 \\
J0436 & S & 1.05 & 0.284 \\
J0452 & S & 0.72 & 0.995 \\
J0708 & C & 1.83 & $<$ 0.10 \\
J1325 & C & 1.39 & $<$ 0.10 \\
J1511 & S & 1.02 & 0.401 \\
J1638 & C & 1.20 & $<$ 0.10 \\
J1937 & C & 3.25 & $<$ 0.10 \\
J2135 & S & 1.09 & 0.213 \\
J2245 & S & 0.94 & 0.566 \\
\hline
\end{tabular}
\end{table}

We further collected the black hole mass from \citet{Chen2018a} and the X-ray flux in 0.2-12.0 keV from the 3XMM-DR8 catalog \citep{Rosen2016}. The X-ray luminosity in 0.2-12.0 keV is calculated using
\begin{equation}
L_X = 4 \pi D_L^2 \cdot F_X, \label{lx}
\end{equation}
where $D_L$ is the luminosity distance estimated by the cosmological redshift. Five sources have a harder X-ray flux and luminosity detection in 14-195 keV by the \textit{Swift} BAT 105-month survey \citep{Oh2018}. These values are listed in Table~\ref{data}.

\begin{table*}
\centering
\caption{The X-ray flux and luminosity, and black hole mass of the 11 NLS1s. \textbf{Columns:} (1) short name, (2) X-ray flux in 0.2-12.0 keV from the 3XMM-DR8 catalog \citep{Rosen2016}, (3) X-ray luminosity in 0.2-12.0 keV derived by Eq.~(\ref{lx}), (4) X-ray flux in 14-195 keV from the Swift-BAT 105-month catalog \citep{Oh2018}, (5) X-ray luminosity in 14-175 keV from the Swift-BAT 105-month catalog \citep{Oh2018}, (6) Black hole mass from \citet{Chen2018a}, (7) X-ray classification, complex (C) or simple (S). \textbf{Notes:} \textit{a.} Other reverberation mapping measurement from the H$\beta$ line gives $M_{BH} = 3.0 \times 10^6 M_{\odot}$ \citep{Wang2001,Du2015}. \textit{b.} Another black hole mass measurement based on the X-ray variability gives $M_{BH} = 2.0 \times 10^7 M_{\odot}$ \citep{Zhou2007}. \textit{c.} Other reverberation mapping measurement gives $M_{BH} = 5.0 \times 10^6 M_{\odot}$ \citep{Zoghbi2010}. \textit{d.} Another black hole mass measurement gives the same $M_{BH} = 1.5 \times 10^7 M_{\odot}$ \citep{Kaspi2000}. \textit{e.} Other black hole mass estimation from the FWHM of H$\beta$ line gives $M_{BH} = 3.0 \times 10^6 M_{\odot}$ \citep{Rodriguez2000,Malizia2008}.}
\label{data}
\begin{tabular}{ccccccc}
\hline
\hline
Short name & $F_{XMM-Newton}$ & $\log_{10} L_{XMM-Newton}$ & $F_{Swift-BAT}$ & $\log_{10} L_{Swift-BAT}$ & $\log_{10} M_{BH}$ & Type \\
- & (10$^{-12}$ erg cm$^{-2}$ s$^{-1}$) & (erg s$^{-1}$) & (10$^{-12}$ erg cm$^{-2}$ s$^{-1}$) & (erg s$^{-1}$) & ($M_{\odot}$) & (C/S) \\
\hline
J0228 & 2.94 & 45.44 & - & - & 7.77 & S \\
J0230 & 34.71 & 43.33 & 11.87 & 42.86 & 6.29 $^a$ & C \\
J0436 & 24.30 & 43.85 & 18.30 & 43.73 & 6.90 & S \\
J0452 & 6.97 & 45.11 & - & - & 6.97 $^b$ & S \\
J0708 & 5.61 & 43.33 & - & - & 6.66 $^c$ & C \\
J1325 & 8.24 & 43.94 & - & - & 7.18 $^d$ & C \\
J1511 & 13.82 & 43.81 & 32.67 & 44.18 & 6.63 & S \\
J1638 & 10.43 & 43.22 & 20.34 & 43.51 & 6.66 & C \\
J1937 & 49.06 & 43.06 & 25.15 & 42.77 & 6.03 $^e$ & C \\
J2135 & 6.18 & 43.74 & - & - & 6.49 & S \\
J2245 & 4.61 & 44.72 & - & - & 7.51 & S \\
\hline
\end{tabular}
\end{table*}

\vspace{10pt}
\textbf{J0228 (IRAS 02262-4110)}

We reproduced the spectrum of J0228 between 2-12 keV with a simple power-law model and obtained a $\chi^2_\nu$ = 1.05. This indicates that a power-law represents the data very well and thus it is classified as a S-NLS1. The \textit{XMM-Newton} EPIC (PN+MOS1+MOS2) spectra of J0228 in the range of 0.2-12.0 keV can be modeled by two power-law components plus an iron line at the rest frame, as seen in \citet{Krumpe2010}.

\vspace{10pt}
\textbf{J0230 (Mrk 1044)}

The spectral fitting of J0230 between 2-12 keV reveals that a simple power-law is not a good representation for the data giving a $\chi^2_\nu$ = 1.76. Indeed, we classify it as a C-NLS1, as it exhibits a relativistic reflection from a high-density accretion disk with the presence of a strong soft X-ray excess below 1.5 keV, Fe K$\alpha$ emission complex at 6-7 keV, and a Compton hump at 15-30 keV in its broad-band (0.3-50.0 keV) spectrum \citep{Mallick2018}. This object had been analyzed by \citet{Mallick2018} in detail. The combined \textit{XMM-Newton} EPIC-PN (0.3−10.0 keV), \textit{Swift} XRT (0.3−6.0 keV), and \textit{NuSTAR} FPMA+FPMB (3.0−50.0 keV) spectra of J0230 can be reproduced by a Galactic absorption, a primary power-law emission, an ionized high-density relativistic disk reflection, a neutral distant reflection, and a Gaussian absorption line. We note that J0230 was classified as a S-NLS1 in \citet{Gallo2006}, whereas we consider it as a C-NLS1. If the flux state scenario is true, this object might have undergone a flux change from normal to low state.

\vspace{10pt}
\textbf{J0436 (Mrk 618)}

A simple power-law modeling in the range of 2-12 keV leads to a $\chi^2_\nu$ = 1.05, indicating that J0436 is classified as a S-NLS1. The \textit{XMM-Newton} EPIC-PN spectrum in 0.2-12.0 keV range can be fitted by a Galactic absorption, a weak intrinsic absorption, a power-law continuum, a blackbody component, and a Fe K$\alpha$ emission line, as reported in \citet{Laha2018}.

\vspace{10pt}
\textbf{J0452 (IRAS 04505-2958)}

We fitted the spectrum of J0452 between 2-12 keV with a simple power-law component reaching a $\chi^2_\nu$ = 0.72, which reveals that J0452 is a S-NLS1. Its \textit{XMM-Newton} EPIC (PN+MOS1+MOS2) spectra in 0.3-10.0 keV can be reproduced by a power-law plus blackbody model or a relativistically blurred photoionized disk reflection model, with additionally a weak intrinsic absorption, an O VII absorption edge, and a Fe K$\alpha$ emission line. Both models give acceptable fits, as analyzed by \citet{Zhou2007}.

\vspace{10pt}
\textbf{J0708 (1H 0707-495)}

J0708 is a well-known C-NLS1s. The spectral fitting in the range of 2-12 keV gives a $\chi^2_\nu$ = 1.83, indicating that a simple power-law model is not a good representation of the data. Indeed, its X-ray spectrum exhibits evident complex features, such as a sharp drop around 7 keV at the rest frame \citep{Boller2002} and/or broad iron K and L emission lines \citep{Fabian2009}, which are usually explained as a relativistically blurred reflection from an iron-rich accretion disk. The \textit{XMM-Newton} EPIC-PN time-integrated spectrum of J0708 in 0.3-10.0 keV had been modeled by \citet{Kara2013} in detail, including components: a Galactic absorption, an intrinsic absorption, a blackbody component, two power-law continua, a relativistically blurred component, and two reflectors irradiated by two power-law components. The hard power-law component is originated from a region close to the center with short time-scale variability, and the soft power-law component is produced in an extended region with long variation. Both power-law continua likely contribute to the reflection.

\vspace{10pt}
\textbf{J1325 (IRAS 13224-3809)}

J1325 is also a C-NLS1s since a simple power-law fitting between 2-12 keV gives a $\chi^2_\nu$ = 1.39 which is not an acceptable fit. Its X-ray spectrum indeed shows a remarkable sharp drop at around 8 keV and a significant ionized Fe L absorption feature at around 1.2 keV \citep{Boller2003}. The \textit{XMM-Newton} EPIC (PN+MOS1+MOS2) time-averaged spectra of J1325 in 0.3-10.0 keV can be reproduced by a Galactic absorption, a power-law continuum, a blackbody component either directly originated from the accretion disk or reprocessed the thermal emission, and two relativistically blurred reflections. Details are analyzed by \citet{Chiang2015}.

\vspace{10pt}
\textbf{J1511 (IRAS 15091-2107)}

A simple power-law is a very good reproduction of the data between 2-12 keV obtaining a $\chi^2_\nu$ = 1.02, which reveals that J1511 is a S-NLS1s. Its combined \textit{XMM-Newton} EPIC-PN (0.3-10.0 keV) and \textit{INTEGRAL} IBIS (20-100 keV) spectra can be modeled by a partially covering absorption, a power-law continuum, and an iron K$\alpha$ emission line, as seen in \citet{Panessa2011}.

\vspace{10pt}
\textbf{J1638 (IRAS 16355-2049)}

The spectral fitting of J1638 gives a $\chi^2_\nu$ = 1.20 using a simple power-law between 2-12 keV. Even though the complex feature is not as evident as in some well-know C-NLS1s, considering the confidence level of the null hypothesis, we classified this source as a C-NLS1. This is in agreement with what was found in \citet{Panessa2011}, who modeled the combined \textit{XMM-Newton} EPIC-PN (0.3-10.0 keV) and \textit{INTEGRAL} IBIS (20-100 keV) spectra of J1638 by a power-law continuum, a soft emission from plasmas, a multi-blackbody component, and two iron K-shell emission lines.

\vspace{10pt}
\textbf{J1937 (IRAS 19348-0619)}

We classify J1937 as a C-NLS1 since a simple power-law can not reproduce the data between 2-12 keV resulting in a very bad spectral fit of $\chi^2_\nu$ = 3.25. Indeed, its broad-band X-ray spectrum (0.3-60.0 keV) exhibits extremely complex features, such as dramatic X-ray variability, strong soft excess below 2 keV, evident relativistic reflection, broad Fe K$\alpha$ line at around 6-7 keV, and Compton hump above 10 keV \citep{Frederick2018}. The combined \textit{XMM-Newton} EPIC-PN (0.3−10.0 keV) and \textit{NuSTAR} FPMA+FPMB (3.0−60.0 keV) time-integrated spectra of J1937 can be modeled by a Galactic absorption, a multi-blackbody accretion disk, and a relativistic reflection model for a lamp post geometry as well as an incident cutoff power-law continuum. Details are analyzed by \citet{Frederick2018}.

\vspace{10pt}
\textbf{J2135 (IRAS F21325-6237)}

\begin{figure}
\centering
\includegraphics[width=\columnwidth]{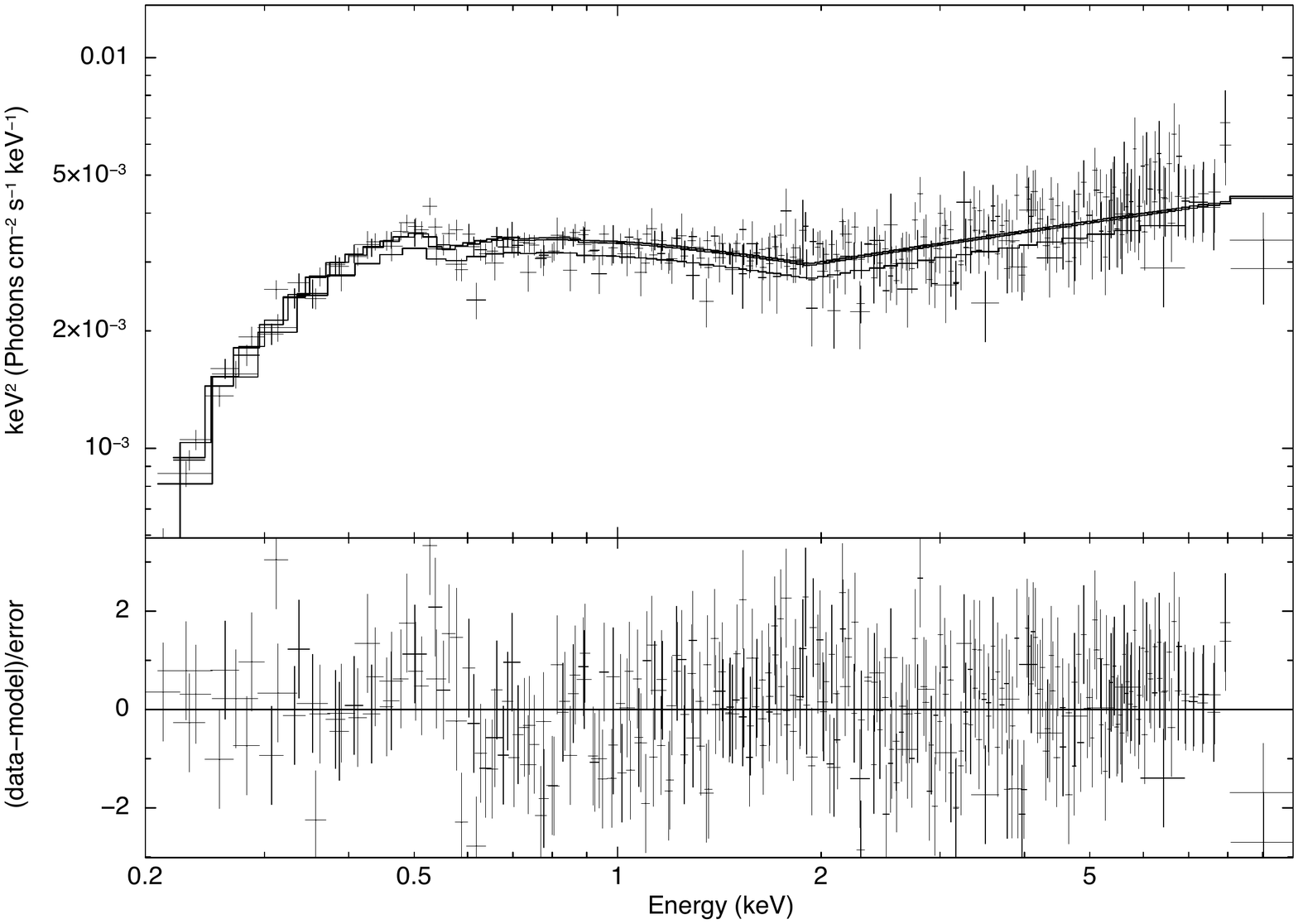}
\caption{The best-fitting model for the X-ray spectrum of J2135 in 0.2-12.0 keV energy band.}
\label{j2135_fit}
\end{figure}

A simple power-law fitting gives a $\chi^2_\nu$ = 1.09 in the range of 2-12 keV, indicating that J2135 is a S-NLS1. Since no X-ray spectrum or modeling was published for this object, we reproduced its \textit{XMM-Newton} spectrum in 0.2-12.0 keV using a broken power-law model, which yields a $\chi^2$/d.o.f. = 354/322 = 1.10, as shown in Table~\ref{j2135_parameters} and Fig.~\ref{j2135_fit}. The broken power-law model gives photon indexes of $\Gamma_1$ = 2.29 and $\Gamma_2$ = 1.75 below and above a break energy of $E_{break}$ = 1.93 keV. No high-energy complexity is shown.

\begin{table}
\centering
\caption{X-ray spectral fitting parameters of J2135. \textbf{Lines:} (1) Galactic hydrogen column density fixed from \citet{HI4PI2016}, (2) the first photon index of the broken power-law, (3) break energy of the broken power-law, (4) the second photon index of the broken power-law, (5) normalization of the broken power-law, (6) cross-calibration constant for MOS1, (7) cross-calibration constant for MOS2, (8) Chi-squared and degree of freedom.}
\label{j2135_parameters}
\begin{tabular}{cc}
\hline
\hline
\textbf{model} & \textbf{wabs * bknpower} \\
\hline
$n_H$ & 3.12 $\times$ 10$^{20}$ cm$^{-2}$ \\
$\Gamma_1$ & 2.29 $\pm$ 0.02 \\
$E_{break}$ & 1.93 $^{+0.20}_{-0.17}$ keV \\
$\Gamma_2$ & 1.75 $^{+0.05}_{-0.06}$ \\
norm & (3.34 $\pm$ 0.09) $\times$ 10$^{-3}$ \\
$c_{MOS1}$ & 1.08 $\pm$ 0.03 \\
$c_{MOS2}$ & 1.09 $\pm$ 0.03 \\
\hline
$\chi^2$/d.o.f. & 354/322 = 1.10 \\
\hline
\end{tabular}
\end{table}

\vspace{10pt}
\textbf{J2245 (IRAS F22423-4707)}

\begin{figure}
\centering
\includegraphics[width=\columnwidth]{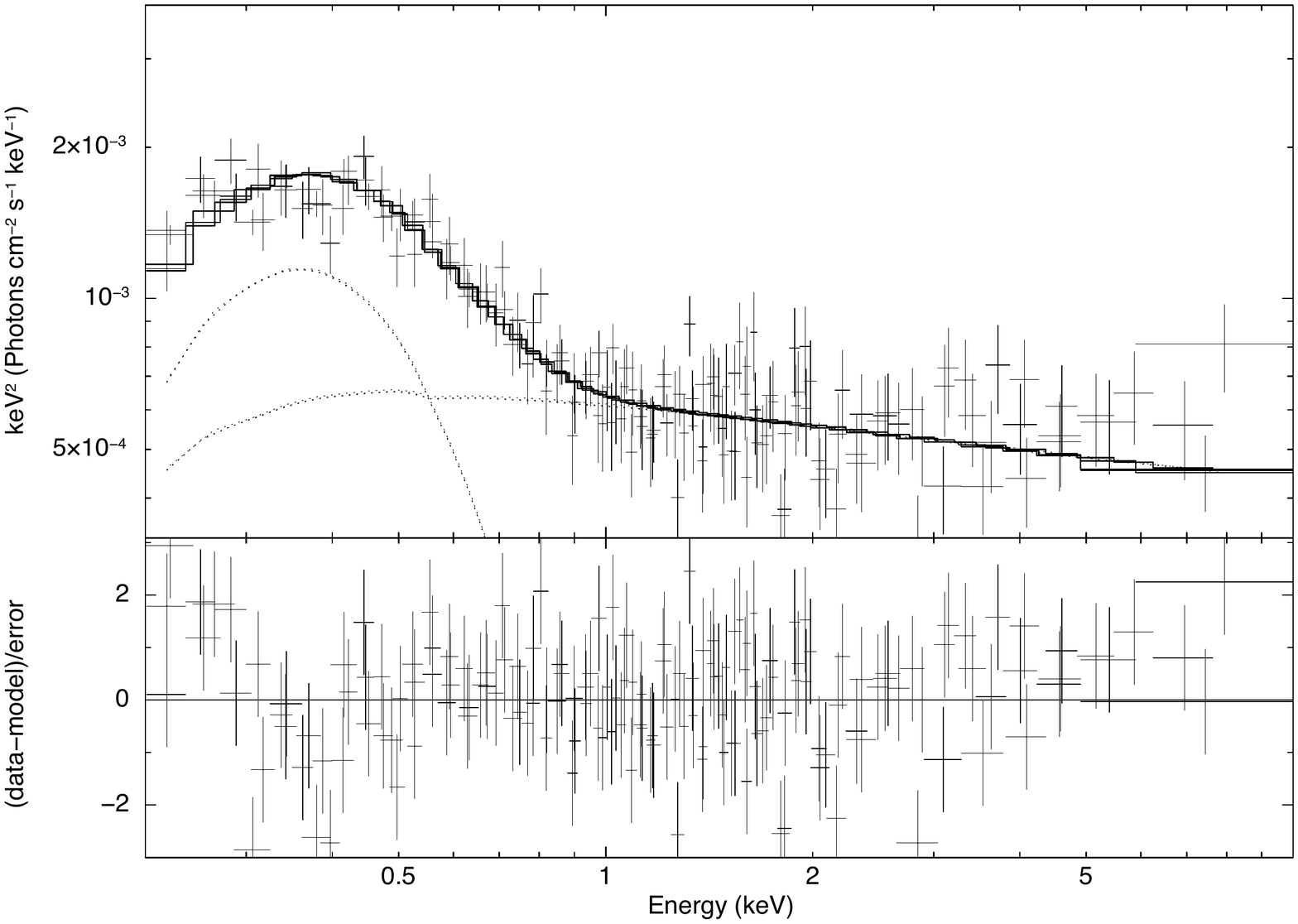}
\caption{The best-fitting model for the X-ray spectrum of J2245 in 0.2-12.0 keV energy band.}
\label{j2245_fit}
\end{figure}

The spectral fitting with a simple power-law between 2-12 keV reaches a $\chi^2_\nu$ = 0.94, suggesting that J2245 is a S-NLS1. This object has never been analyzed in the literature. Its \textit{XMM-Newton} spectrum in the range of 0.2-12.0 keV can be reproduced by a blackbody plus power-law model, yielding a blackbody temperature of $kT$ = 0.10 keV and a photon index of $\Gamma$ = 2.16. This model has a $\chi^2$/d.o.f. = 189.4/157 = 1.21, as seen in Table~\ref{j2245_parameters} and Fig.~\ref{j2245_fit}. No complex feature is evident in the high-energy part of the spectrum.

\begin{table}
\centering
\caption{X-ray spectral fitting parameters of J2245. \textbf{Lines:} (1) Galactic hydrogen column density fixed from \citet{HI4PI2016}, (2) temperature of the blackbody, (3) normalization of the blackbody, (4) photon index of the power-law, (5) normalization of the power-law, (6) cross-calibration constant for MOS1, (7) cross-calibration constant for MOS2, (8) Chi-squared and degree of freedom.}
\label{j2245_parameters}
\begin{tabular}{cc}
\hline
\hline
\textbf{model} & \textbf{wabs * (zbbody + zpowerlw)} \\
\hline
$n_H$ & 0.94 $\times$ 10$^{20}$ cm$^{-2}$ \\
$kT$ & 0.10 $\pm$ 0.00 keV \\
norm & (5.30 $\pm$ 0.41) $\times$ 10$^{-5}$ \\
$\Gamma$ & 2.16 $\pm$ 0.07 \\
norm & (9.35 $^{+0.58}_{-0.57}$) $\times$ 10$^{-4}$ \\
$c_{MOS1}$ & 0.99 $\pm$ 0.04 \\
$c_{MOS2}$ & 0.99 $\pm$ 0.04 \\
\hline
$\chi^2$/d.o.f. & 189.4/157 = 1.21 \\
\hline
\end{tabular}
\end{table}

\section{Discussion}

\subsection{The X-ray complexity}

The X-ray spectral complexity is due to the complex Fe K-shell features, which produce strong emission and absorption features around 6-8 keV in the rest frame of the galaxy appearing as a broad emission line or a sharp spectral drop. These spectral features are described as arising from extremely smeared and relativistic reflection from ionized material in the inner parts of the accretion disk surrounding a rapidly spinning black hole, with the additional requirement of strongly super-solar iron abundance \citep{Fabian2013}. An alternative explanation is that the complex Fe K-shell features are caused by a clumpy wind from the inner accretion disk, without requiring any extreme relativistic smearing or super-solar iron abundance \citep{Hagino2016, Jin2017b}. In this scenario, our line of sight crosses the acceleration region where the outflow or wind is launched. The wind leads to the iron emission and absorption lines covering a wide range of velocities, thereby resulting in a broad emission line or a deep absorption drop around 7 keV.

\subsection{The flux state scenario}

\begin{figure}
\centering
\includegraphics[width=\columnwidth]{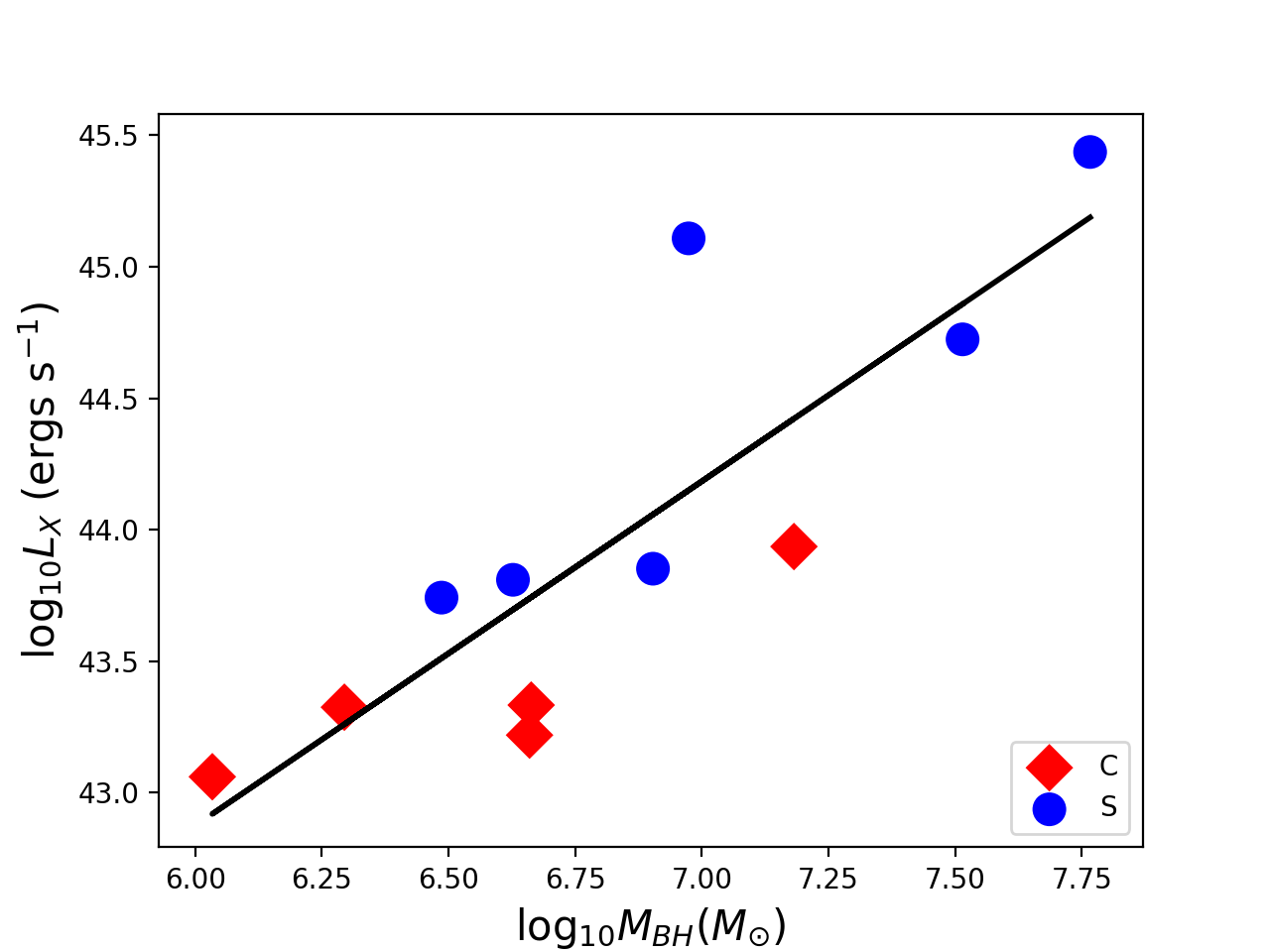}
\caption{The linear correlation (black line) between the black hole mass and the X-ray luminosity for C-NLS1s (red diamonds) and S-NLS1s (blue circles).}
\label{m+l_s+c}
\end{figure}

We plotted the relation between the black hole mass $\log_{10} M_{BH}$ and the X-ray luminosity $\log_{10} L_X$, as shown in Fig.~\ref{m+l_s+c}. It is worth noting that there may be a linear correlation as found by \citet{Bianchi2009, Jarvela2015},
\begin{equation}
\log_{10} L_X = (1.3 \pm 0.4) \times \log_{10} M_{BH} + (35.0 \pm 9.7),
\end{equation}
even though it is not so tight with values of $r$ = 0.85 and $p$ = 1.05 $\times$ 10$^{-3}$. Moreover, C-NLS1s and S-NLS1s seem to occupy the low and high $M_{BH} - L_X$ regions of the plot respectively.

We used a two-sample Kolmogorov-Smirnov (K-S) test to examine the distributions of black hole mass and X-ray luminosity of the C-NLS1 and S-NLS1 subsamples. The null hypothesis is that two distributions originate from the same population of sources. We applied the rejection of the null hypothesis at a 90$\%$ confidence level corresponding to a value of $p \leq 0.10$, which is at the same confidence level of the classification of C-NLS1s and S-NLS1s. The K-S test on the X-ray luminosity ($p$ = 0.026) suggests that C-NLS1s and S-NLS1s do not originate from the same population. Instead, the K-S test on the black hole mass ($p$ = 0.454) suggests that C-NLS1s and S-NLS1s are the same population.

This result seems to be in agreement with the conclusion by \citet{Gallo2006}, that is C-NLS1s and S-NLS1s represent sources from the same population but in a low and normal X-ray flux state respectively. According to our results, in fact, C-NLS1s are systematically at a lower luminosity from both \textit{XMM-Newton} and \textit{Swift}-BAT detections, thus the high-energy complexity in the X-ray spectra is visible. If the luminosity becomes higher, the X-ray spectral complexity is less prominent due to the bright continuum overwhelming the emission lines \citep{Foschini2012}, therefore appearing as S-NLS1s. If this scenario is correct, the classification of S-NLS1s or C-NLS1s is a transient condition, strongly dependent on the activity state of each source.

\subsection{The outflow scenario}

\begin{figure}
\centering
\includegraphics[width=\columnwidth]{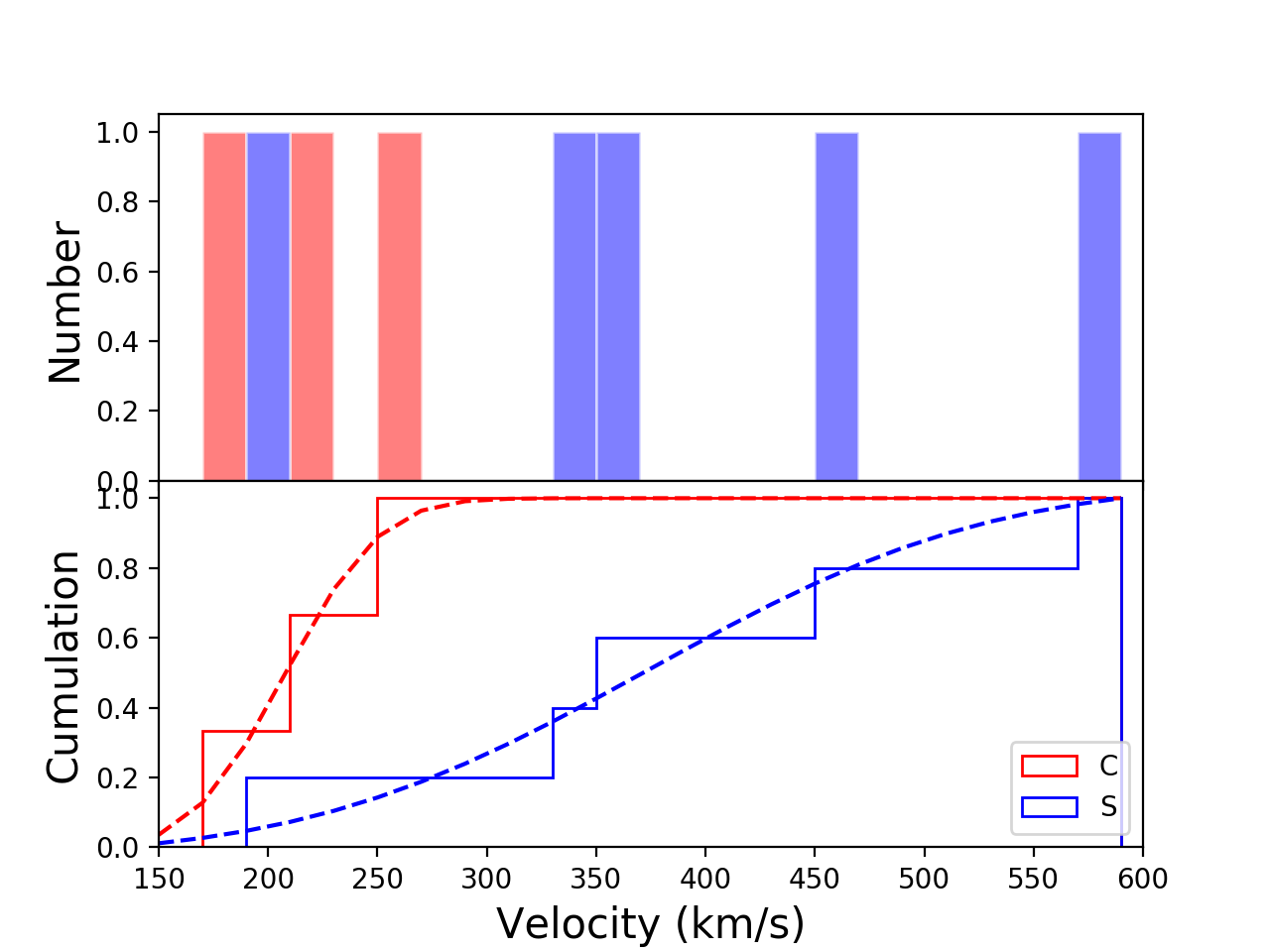}
\caption{The velocity distribution of the blue wing in [O III] $\lambda$5007 line for C-NLS1s (red) and S-NLS1s (blue) using a 20 bin width.}
\label{velocity}
\end{figure}

To study the [O III] outflow properties, we plotted the velocity distribution of the blue wings which might be associated with the wind interacting with the NLR material and appearing as the gas outflows \citep{Berton2016a}, as displayed in Fig.~\ref{velocity}. We found that such outflows from the acceleration region are present in both C-NLS1s and S-NLS1s. Indeed, NLS1s with a high accretion rate are very likely to power a wind by means of the radiation pressure coming from the accretion disk \citep{Proga2000, Parker2017}.

As before, we performed a K-S test to verify if the blue wing velocity distribution of the C-NLS1 and S-NLS1 subsamples are the same. We applied the rejection of the null hypothesis that two distributions originate from the same population at a 90$\%$ confidence level with a value of $p \leq 0.10$. The result of the K-S test ($p$ = 0.085) suggests that the blue wing velocity in C-NLS1s and S-NLS1s have different distributions, even though not in a high confidence level due to the small sample and the large uncertainty of the velocity measurement. The wind in S-NLS1s tends to be faster than that in C-NLS1s.

A possible explanation is that the low velocity of the wind observed in C-NLS1s corresponds to a low energy, which is not sufficient to blow away the ionized material. Therefore, such material remains close to the central engine where it can produce the X-ray complexity. On the other hand, in S-NLS1s the wind velocity and energy are significantly higher, thus it is able to clean the ionized material in its path and cause the lack of complex feature in the X-ray spectra. 

\subsection{The inclination scenario}

As mentioned above, the X-ray spectral complexity might be an inclination effect \citep{Jin2017a, Jin2017b}. The outflows observed as blue wings in the [O III] lines are mostly directed along the system axis perpendicularly to the accretion disk. If the source is seen at a low inclination, the outflows will be directed toward the observers, and the blueshift will be large. Furthermore, along the system axis the line of sight is free of obscuration because all the ionized material is blown away by such a strong wind. In fact, as shown by a magnetohydrodynamic simulation, outflows along the line of sight (i $<$ 15$^\circ$) are those with higher speeds $\sim$ (0.3-0.4)c \citep{Yuan2015}. At a larger inclination, instead, the blueshifted components of the [O III] lines will be less prominent, and as shown by \citet{Jin2017b}, obscuring material will be present along the line of sight, thus producing the observed complexity.

Therefore, S-NLS1s may be sources viewed at small inclination angles, and their blue wings are faster because they are directed toward the observers. Since the line of sight is clear from obscuration, what we observe is the X-ray spectrum of the central engine. Conversely, C-NLS1s are sources viewed at large inclination angles, in which the blueshift of the wings is smaller, and the presence of ionized material leads to the X-ray spectral complexity.

This scenario has important implications on the nature of NLS1s. If this hypothesis is correct, the X-ray complexity could be used as an inclination indicator for these sources. Some authors proposed that the narrow permitted lines typical of NLS1s may be the product of a disk-like broad-line region (BLR) observed pole-on \citep{Decarli2008}. In this framework, if C-NLS1s are objects seen at a large inclination with a flat BLR, their permitted lines should be broader than those of S-NLS1s. However, such difference has not been observed \citep{Vietri2018}. This indicates that either C-NLS1s are not viewed at larger angles, or the BLR in NLS1s is not flattened. The latter hypothesis is also in agreement with the physical interpretation of the Lorentzian line profiles observed in NLS1s \citep{Kollatschny2013}.

However, it is worth noting that this scenario does not account for the difference we observed in the X-ray luminosity of simple and complex sources. As mentioned above, C-NLS1s have on average a lower luminosity with respect to S-NLS1s, and this result seems to be easily explainable in the flux state scenario. Further observations on a larger sample are needed to confirm or disprove these findings.

\section{Summary}

In this work, we analyzed 11 NLS1s in the 6dFGS sample with both optical and X-ray spectroscopic observations. Combining the optical and X-ray spectral properties, we propose a possible correlation between [O III] line asymmetry and X-ray spectral complexity in NLS1s. 

The outflow or wind is probably formed via radiation pressure from the inner accretion disk of NLS1s. Its interaction with the material in the inner NLR may be seen in terms of blue wing velocity in the [O III] emission lines. In C-NLS1s only weak wind effects are measured, the X-ray spectral complexity can be well explained by the ionized material in the wind leading to the complex features. On the contrary, the wind effects detected in S-NLS1s are strong and very likely to blow away the ionization material, therefore resulting in the X-ray spectral simplicity.

We also suggest that this correlation is due to an inclination effect. S-NLS1s may be sources viewed at small inclination angles, in which the X-ray spectrum is not obscured by any intervening medium. Since the outflows are mostly directed along the system axis, the bulk of the gas is moving toward the observers, and the blueshift of the emission lines is maximum. C-NLS1s instead may be sources viewed at large inclination angles, where the ionized material is still present, and the blueshift effect is less prominent.

Finally, we remark that these results should be taken with caution because of the limited number of NLS1s used in this analysis and the large uncertainty of the blue wing velocity measurements. Further study on a larger sample and higher S/N ratio spectra will be necessary to unveil the peculiar X-ray properties of NLS1s.

\section*{Acknowledgements}

We thank the anonymous referee for suggestions leading to the improvement of this work. JHFan's work is partially supported by the National Natural Science Foundation of China (NSFC 11733001, NSFC U1531245) and Natural Science Foundation of Guangdong Province (2017A030313011). This research is based on observations obtained with \textit{XMM-Newton}, an ESA science mission with instruments and contributions directly funded by ESA Member States and NASA. This project would not have been possible without ignoring the advice of far more people than can be mentioned here. We would like to thank all our colleagues for their suggestions, bug reports, and (especially) source code. The GSFC X-ray astronomy group are particularly thanked for their patience exhibited while functioning as the beta test site for new releases. The initial development of XSPEC was funded by a Royal Society grant to Prof. Andy Fabian and subsequent development was partially funded by the European Space Agency's EXOSAT project and is now funded by the HEASARC at NASA/GSFC. This paper is based on the observations collected with the 2.5-meter Du Pont telescope and the 6.5-meter Walter Baade telescope operated by the Las Campanas Observatory (LCO) in Chile, the 1.22-meter Galileo telescope operated by the Asiago Astrophysical Observatory in Italy, and the 2.54-meter Isaac Newton telescope operated by the Roque de los Muchachos Observatory in Spain. This research has made use of the NASA/IPAC Extragalactic Database (NED), which is operated by the Jet Propulsion Laboratory, California Institute of Technology, under contract with the National Aeronautics and Space Administration.

\bibliographystyle{mnras}
\bibliography{bibliography}

\end{document}